\documentstyle[twoside,epsf,hyperref]{article}
\input ibvs3.sty
\begin{document}
\IBVSheadDOI{63}{6257}{19 January 2019}

\IBVStitletl{UU A{\lowercase{qr}} -- No superhumps but variations}{on the
time scale of days}

\IBVSauth{Bruch, Albert}

\IBVSinst{Laborat\'orio Nacional de Astrof\'{\i}sica, Rua Estados Unidos, 154, CEP 37504-364, Itajub\'a -- MG, Brazil}

\SIMBADobj{UU Aqr}
\GCVSobj{UU Aqr}
\IBVSkey{photometry}

\IBVSabs{Recently, brightness variations occurring on {\it twice} the 
accretion disk precession period in the old nova and permanent superhump
system V603~Aql have been observed by Bruch \& Cook (2018). In an attempt
to detect a similar effect in other cataclysmic variables reported to
contain permanent superhumps the novalike variable UU~Aqr was observed
during 11 nights in September 2018. While no traces of superhumps were
seen in the data, rendering the quest for variations related to the disk
precession period obsolete, the system exhibits regular variations with a
period of $\sim$4 days.}

\begintext

The light curves of some cataclysmic variables exhibit photometric 
variations, termed superhumps, with periods slightly longer than their 
orbital periods. They are thought to be caused by stresses induced by the
periodic passage of the secondary star close to the extended part of the
accretion disk which in these cases is not circular but elliptically
deformed (Whitehurst, 1988). The period is longer than the orbital period 
because the major axis of the accretion disk precesses. An alternative
model is promoted by Smak (2009): the irradiation of the secondary star
by the primary component varies  because of rotating non-axisymmetric vertical 
structures in the accretion disk, leading to a modulation of the mass
transfer rate and in consequence to variable dissipation of kinetic energy.
The superhump phenomenon occurs always during
supermaxima of the short-period dwarf novae of SU~UMa subtype. However,
some novalike variables and old nova also exhibit superhumps (see, e.g.,
Patterson, 1999). (Although these are normally termed ``permanent 
superhumpers'', superhumps in these systems may not always be that permanent!)

One of them is the old nova V603~Aql which has an orbital period of
$P_{\rm orb} = 3.32$~h and a well established (albeit slightly variable) 
superhump period of $P_{\rm SH} \approx 3.5$~h. Recently, Bruch \& Cook (2018) 
found an additional period in the light curve of V603~Aql which is related 
to the beat between $P_{\rm orb}$ and $P_{\rm SH}$, confirming marginal evidence
for this phenomenon presented earlier by Suleimanov et al.\ (2004). Some
other permanent superhump systems with limited evidence for a similar 
behaviour are listed by Yang et al.\ (2017). On the other hand, in SU~UMa
type dwarf novae with high orbital inclination variations of the system
brightness on the beat period are common (Smak, 2009; 2013) and can readily
be explained in Smak's model by the non-axisymmetric structures in the outer
disk. As confirmed observationally by Smak (2009) such modulation should 
therefore not be seen in low inclination systems. Consequently, the beat
period related variations seen in V603~Aql can not be explained within Smak's 
model because the orbital inclination
of $13^{\rm o} \pm 2^{\rm o}$ (Arenas et al.\ 2000) is far to low. 
Moreover, quite intriguingly and in contrast to the 
finding of Suleimanov et al.\ (2004) the period observed very 
clearly by Bruch \& Cook (2018) is not equal to the beat between $P_{\rm orb}$ 
and $P_{\rm SH}$ and thus the precession period, $P_{\rm prec}$, of the accretion 
disk, but exactly twice this value. While there is no obvious reason why the 
system brightness should change with the precession period in this low
inclination system a modulation with $2 \times P_{\rm prec}$ is even 
more mysterious.

In an attempt to verify if similar variations related to $P_{\rm prec}$ occur in
other systems exhibiting permanent superhumps as a first step towards an 
understanding, I observed a series of light curves of
the novalike cataclysmic variable UU~Aqr. This is an eclipsing system
with an orbital period of 0.16580429 days ($\approx 3^{\rm h} 56^{\rm m}$)
(Baptista et al. 1994). Patterson et al.\ (2005) observed a strong superhump
in 2000 with a period of $4^{\rm h} 12^{\rm m}$. But note that in 1998 their
observations yielded only marginal evidence for superhump-like variations.
The orbital and superhump periods imply a precession period for the 
accretion disk of 3.12 days.

I used the 60\,cm Boller \& Chivens telescope of the Observat\'orio do
Pico dos Dias, Brazil, to observe UU~Aqr in 11 nights between 2018, 
September 6 and 17. Light curves in unfiltered light spanning more than
8 hours in most of the nights were obtained at a time resolution of 5~sec.
Synthetic aperture photometry of UU~Aqr was performed on the original 
images (using a blue-sensitive IKon-L936-BEX2-DD CCD) 
after bias subtraction and 
flat-fielding, employing the MIRA software system (Bruch 1993). Magnitudes were
measured relative to the primary comparison star \#05 (Henden \&
Honeycutt, 1995; $V = 13.804$). For cataclysmic
variables the throughput of the instrumentation corresponds roughly to $V$
(Bruch, 2018). The light curves are shown in Fig.~1 where
the time and magnitude scales are the same for all frames. Apart from
eclipses they are characterized by rather strong flickering and modest
variations on the time scale of hours which, however, exhibit no obvious
regularity. 

\IBVSfig{14cm}{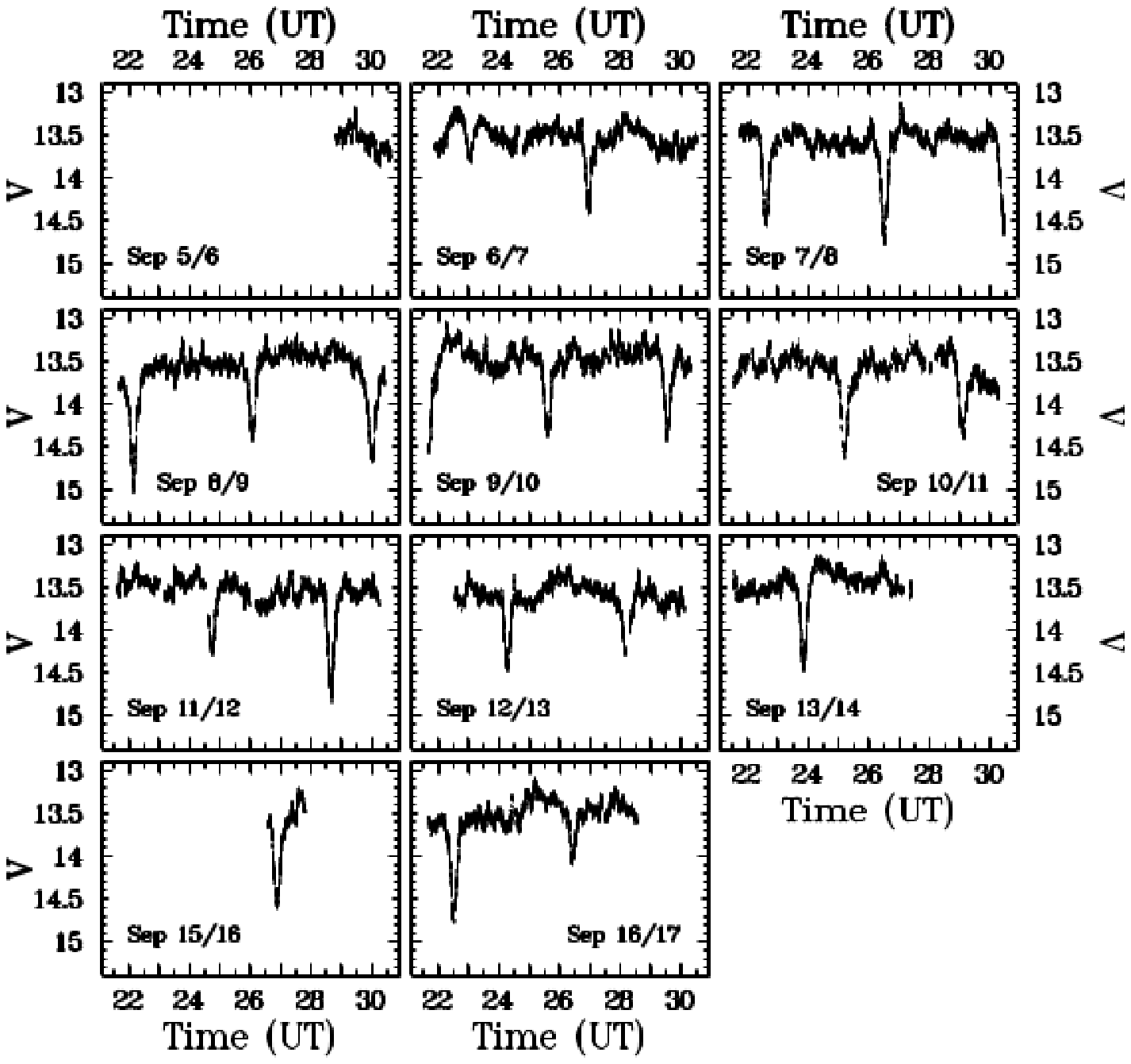}{Light curves of UU~Aqr observed in 11 nights
in 2018 September, all drawn on the same time and magnitude scale.}
\IBVSfigKey{6257-f1.eps}{UU Aqr}{light curve}

As an aside I draw attention to the strong variability of the
eclipse depth which occurs even during subsequent cycles. This is particularly
striking on September 6/7, where the eclipse close to UT 23~h hardly stands
out in the light curve. Apparently, the secondary star in UU~Aqr only 
partially covers the brighter parts of the primary and variations in the
brightness of the central region of the accretion disk can strongly modulate
the eclipse depth. 

Turning now to the main purpose of the observations of UU~Aqr, i.e., the
investigation of a possible relationship between orbital, superhump and
accretion disk precession period, I first masked the eclipses
because they would dominate any period search algorithm. In
order to remove any light travel time effects in the solar system, time
was then transformed into barycentric Julian Dates on the Barycentric
Dynamical Time scale, using the online tool of Eastman et al.\ (2010).
Thereafter, all light curves were combined into a single data set. The
result is shown in the upper frame of Fig.~2.

\IBVSfig{14cm}{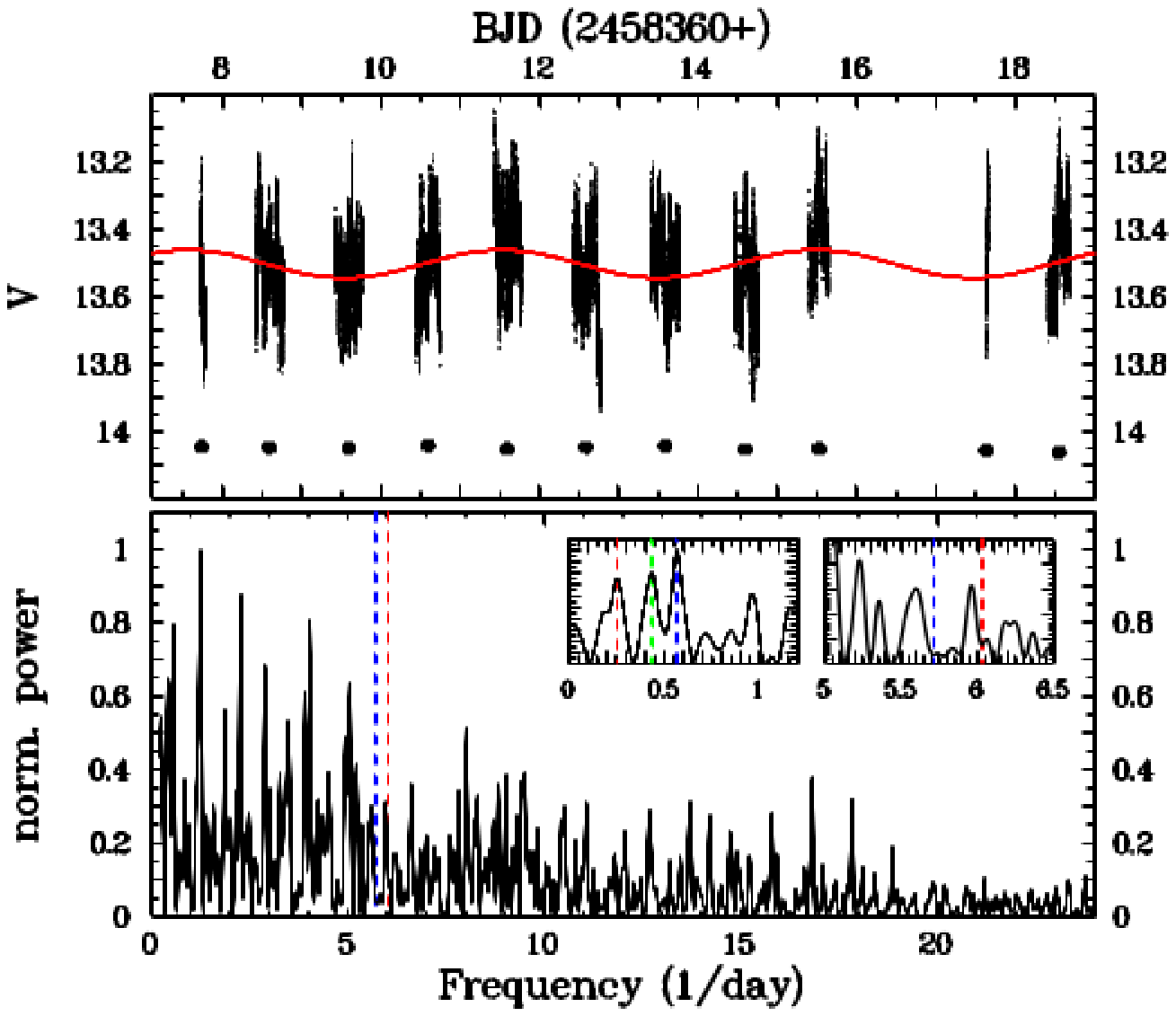}{{\it Top:} The combined light curves of
UU~Aqr of 2018, September, after removal of eclipses. The dots below the
light curves represent the nightly averages of the magnitude difference
between the primary comparison star and a check star.  {\it Bottom:} 
Lomb-Scargle periodogram of the light curves shown in the upper frame.
The broken vertical lines indicate the orbital (red) and previously
observed superhump frequency (blue). In inserts show blown up versions
of a small part of the periodogram around the orbital and superhump
frequencies (right) and of the low frequency part of the spectrum (left),
with some prominent peaks marked by vertical lines.}
\IBVSfigKey{6257-f2.eps}{UU Aqr}{periodogram}

A power spectrum of the combined light curve was calculated using the
Lomb-Scargle algorithm (Lomb 1976; Scargle 1982). 
The lower frame of Fig.~2 contains
the resulting periodogram. Several peaks are visible, but none of them
stands out among the others. Moreover, the power spectra of subsets of
all data do not contain significant signals at the same frequencies. Therefore, 
none of the peaks in the power spectra of the combined data indicates
a stable period in UU~Aqr. In particular, neither the orbital period nor
the previously observed superhump period manifest themselves in the power
spectrum. The respective frequencies are marked by the blue and red
 vertical lines in the figure, respectively. The right hand inset contains a 
blown-up version of a small frequency range around $1/P_{\rm orb}$ and 
$1/P_{\rm SH}$. It must therefore be concluded that the superhumps seen in 
2000 by Patterson et al.\ (2005) have vanished. The absence of any signal at
$1/P_{\rm orb}$ also indicates that apart from the eclipses UU~Aqr does not
exhibit orbital variations such as a an orbital hump -- often seen in
cataclysmic variables -- caused by a hot spot at the location where the 
matter transferred from the secondary star hits the accretion disk. 

The absence of superhumps turns the quest for variations related to the
beat between orbit and superhump obsolete. Even so, the combined light
curve (upper frame of Fig.~2) contains systematic night-to-night variations
which apparently are not random. Their significance can be assessed through
the behaviour of the comparison and check stars. Since the nightly averages 
of the magnitude differences between the primary comparison
star and 4 check stars revealed a slight (amplitude $\le 0.02$~mag)
systematic variation of the former, well approximated by a third order
polynomial, a corresponding correction has been
applied to all light curves. The comparison -- check star light curves then 
becomes virtually flat. One of them is plotted (shifted in magnitude by an 
arbitrary constant) below the UU~Aqr light curve in Fig.~2.

The long-term variations should reveal themselves also at the low frequency
end of the power spectrum which is plotted at an enlarged scale in the left-hand 
inset of the figure. The strongest peaks are marked by coloured 
vertical lines and correspond to periods of 3.966, 2.304 and 1.773 days.
There is no obvious mutual relationship between these values or with the
orbital or superhump period. Moreover, it is not straightforward to assess
their statistical significance.
Least squares sine fits with these periods yield half amplitudes
of 0.042, 0.045, and 0.051 mag, respectively. The shorter periods do not
reveal themselves intuitively to the eye. They are also not seen
clearly in the power spectra of subsets of the data. However, trusting in the
high capability of the human brain for pattern recognition, the reality of the 
$\sim$4 day period (red curve in the figure) is more convincing. 
While the data may not be sufficient to claim that this variation is
really periodic and stable over time, it occurs on the same order of magnitude 
of the expected disk precession period if the superhumps were present. However,
this may be a mere coincidence. 

Concluding, I remark that in September 2018 UU~Aqr did not exhibit 
superhumps and that these are thus not a permanent feature in the light
curve of the system. This renders impossible the original purpose of this
work, i.e., the investigation of brightness variations related to the
precession period between the orbit and superhump periods. Nevertheless,
UU~Aqr exhibits systematic brightness variations on similar time scales,
although the data do not permit a definite claim for their stability
and repeatability.

\vskip 1cm

{\it Acknowledgements:} This work is exclusively based on observations
obtained at the Observat\'orio do Pico dos Dias, maintained by the
Laborat\'orio Nacional de Astrof\'{\i}sica, a branch of the Minist\'erio
da Ci\^encia, Tecnologia, Inova\c{c}\~ao e Comunica\c{c}\~oes da
Rep\'ublica Federativa do Brasil.

\references 

Arenas, J., Catal\'an, M.S., Augusteijn, T., Retter, A., 2000, 
{\it MNRAS}, {\bf 311}, 135
\BIBCODE{2000MNRAS.311..135A} \DOI{10.1046/j.1365-8711.2000.03061.x}

Baptista, R., Steiner, J.E., Cieslinski, D., 1994, {\it ApJ}, {\bf 433}, 332
\BIBCODE{1994ApJ...433..332B} \DOI{10.1086/174648}

Bruch, A. 1993, ``MIRA: A Reference Guide'', Astron.\ Inst.\ Univ.\ M\"unster

Bruch, A., 2018, {\it New Astr.}, {\bf 58}, 53
\BIBCODE{2018NewA...58...53B} \DOI{10.1016/j.newast.2017.07.007}

Bruch, A., Cook, L.M., 2018, {\it New Astr.}, {\bf 63}, 1
\BIBCODE{2018NewA...63....1B} \DOI{10.1016/j.newast.2018.02.002}

Eastman, J., Siverd, R., Gaudi, B.S., 2010, {\it PASP}, {\bf 122}, 935
\BIBCODE{2010PASP..122..935E} \DOI{10.1086/655938}

Henden, A.A., Honeycutt, R.K., 1995, {\it PASP}, {\bf 107}, 324
\BIBCODE{1995PASP..107..324H} \DOI{10.1086/133557}

Lomb, N.R., 1976, {\it Ap\&SS}, {\bf 39}, 447
\BIBCODE{1976Ap&SS..39..447L} \DOI{10.1007/BF00648343}

Patterson, J., 1999, in S.\ Mineshige \& C.\ Wheeler (eds.) ``Disk 
Instabilities in Close Binary Systems'', Universal Academy Press, Tokyo, p.\ 61
\BIBCODE{1999dicb.conf...61P} 

Patterson, J., Kemp, J., Harvey, D.A., et al., 2005, 
{\it PASP}, {\bf 117}, 1204 
\BIBCODE{2005PASP..117.1204P} \DOI{10.1086/447771}

Scargle, J.D., 1982, {\it ApJ}, {\bf 263}, 853
\BIBCODE{1982ApJ...263..835S} \DOI{10.1086/160554}

Smak, J., 2009, {\it Acta Astr.}, {\bf 59}, 121
\BIBCODE{2009AcA....59..121S} 

Smak, J., 2013, {\it Acta Astr.}, {\bf 63}, 369
\BIBCODE{2013AcA....63..369S} 

Suleimanov, V., Bikmaev, I., Belyakov, K., et al., 2004, 
{\it Astron.\ Lett.}, {\bf 30}, 615
\BIBCODE{2004AstL...30..614S} \DOI{10.1134/1.1795950}

Whitehurst, R., 1988, {\it MNRAS}, {\bf 232}, 35
\BIBCODE{1988MNRAS.231...35W} \DOI{10.1093/mnras/232.1.35}

Yang, M.T.-C., Chou, Y., Ngeow, C.-C., et al., 2017, {\it PASP}, {\bf 129}, 4202
\BIBCODE{2017PASP..129i4202Y} /DOI{10.1088/1538-3873/aa7a99}

\endreferences

\end{document}